# Mathematical Analysis and Computational Integration of Massive Heterogeneous Data from the Human Retina

Arash Sangari[1], Adel Ardalan[1], Larry Lambe[2], Hamid Eghbalnia[3] and Amir H. Assadi[4]

[1] Department of Electrical and Computer Engineering, University of Wisconsin, USA
[2] Multidisciplinary Software Systems Research Corporation (MSSRC), USA
[3] Department of Molecular and Cellular Physiology, University of Cincinnati, USA
[4] Department of Mathematics, University of Wisconsin, USA

**Abstract**. Modern epidemiology integrates knowledge from heterogeneous collections of data consisting of numerical, descriptive and imaging. Large-scale epidemiological studies use sophisticated statistical analysis, mathematical models using differential equations and versatile analytic tools that handle numerical data. In contrast, knowledge extraction from images and descriptive information in the form of text and diagrams remain a challenge for most fields, in particular, for diseases of the eye. In this article we provide a roadmap towards extraction of knowledge from text and images with focus on forthcoming applications to epidemiological investigation of retinal diseases, especially from existing massive heterogeneous collections of data distributed around the globe.

## 1   Introduction

In epidemiological studies of retinal diseases, one encounters the problem of extracting knowledge from heterogeneous collections of data consisting of numerical, descriptive and imaging. Large scale epidemiological studies use sophisticated statistical analysis, differential equations and related versatile mathematical tools that are developed for numerical data. In contrast, knowledge extraction from retina images and descriptive information in the form of text and diagrams remain scarcely developed. This article provides a roadmap towards epidemiology of diseases of the eye using extraction of knowledge from text, images and numerical data, and heterogeneous data fusion.

In the first part of the article an outline of knowledge extraction from massive text data are discussed. In the second part, we address a general technique for systematic knowledge from very massive retinal image collections. The mathematical tools and concepts are adapted from the PDE based image analysis, which uses advanced numerical and symbolic computation libraries. The algorithms and the ensuing codes are particularly designed for very large industrial-grade projects.

To bring such sophisticated machinery to bear results on retinal image data requires transformation of the raw images into black and white (or gray scale). The simplified images are preprocessed to delineate the anatomical details of the vasculature. Morphometric invariants of the brain structures encountered in such images are, then, extracted in the form of diameter of blood vessels. A weighted graph to capture a combinatorial organization of vessels, branching and notable curvilinear features further a branched tubular surface encodes the endothelial cells that cover the inner part of blood vessels and the whole micro-vessel structures.

The above mentioned data are hierarchically organized to separate distinct phenotypic traits that arise at a particular scale with the appropriate resolution. Finally, the hierarchical graph that encodes combinatorial – quantitative – geometric phenotypic traits could be quantified for massive data analysis using graph theoretic invariants such as the spectrum of Laplace and other operators on graphs. In short, image data provides data structures that quantify morphological traits according to hierarchy resolution and combinatorial features.

## 2   Using Information Extraction Techniques to Enhance Systems Biology

As a result of exploding volume of scientific publications which emerge each day, it's becoming more and more intractable to read and consume the knowledge embedded in the scientific community in an integrated, meaningful and efficient manner. Among the main difficulties is the productivity and utilization of the manuscripts and other types of written material, like descriptive data about the disease or the health status of the patient, for professionals in different fields of study who do not want and probably cannot spend a lifetime reading every minor detail of all the disciplines they deal with in interdisciplinary research activities. In order to be able to use the knowledge and also to communicate with the experts in the field, it is essential to be able to have a general overview of the important entities and their relationships. As an example, biological computation specialists need to understand the roles of different genes and their relationships in order to be able to perform meaningful computations and analysis. Another example is to "understand" the notes and descriptions about patients' state of health scribed during physical examinations.

Information extraction (IE) is among the most promising emerging technique which, in the era of supercomputing and computational clouds, enables fast analysis of drastically huge bulks of documents. IE specialists nowadays think of doing complicated machine learning tasks on the whole web [1] (the size of the indexed Web is estimated to be about 46 billion pages as of November 2011 [2]). This ambitious goal has turned IE techniques into powerful and strong tools which could be utilized to help academic community reducing the time-of-flight for getting into useful and productive research activities. The goal of this phase is to incorporate different large-scale, high-throughput, flexible IE techniques to enhance our problem solving skills and leverage our communications with collaborating professionals in genomic sciences, health care and neuroscience to establish fruitful research projects and to contribute in building a framework for translating the extensive amount of knowledge (and wisdom) buried inside scientific publications descriptive healthcare records, even for decades, to be used in action by large number of researchers and students.

IE usually starts with digesting the free text into a structured form using various natural language processing (NLP) methods [3]. These methods are built upon different sources of information, ranging from linguistic grammatical structures to statistical properties of various types of written and spoken corpora. The main tasks are illustrated in the following figure.

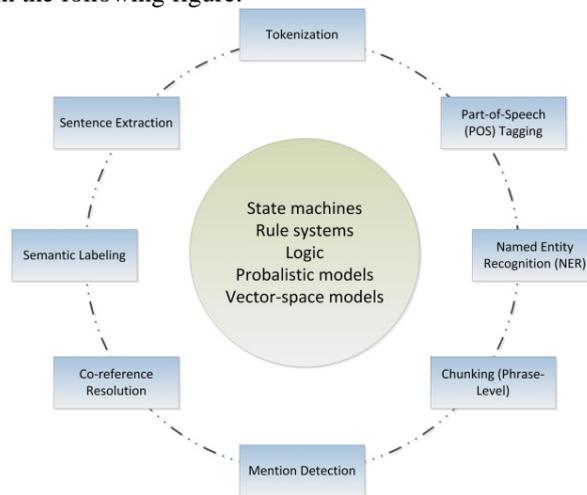

**Fig. 1.** Components of Natural Language Analysis.

A rich literature and set of tools is developed in the area of data integration which tries to unify different data sources in terms of structure and semantic [4]. Different machine learning and statistical analysis tools have been tested and turned out to be useful in different contexts. The more general framework of information fusion deals with making decisions based on heterogeneous sources of information, ranging from numerical data to free text to imagery data. Graphical probabilistic models have provided a flexible and powerful framework to fuse data and information. For example, Markov logic networks (MLN) [5] have been used successfully to extract information from various sources of textual information. As one of the most basic tools in logic, predicate logic and calculus have played a major role in knowledge representation since their formal introduction. In its simplest form, first-order predicate logic is able to encapsulate knowledge from heterogeneous sources, such as the ones mentioned above, and thanks to new tools that have been developed recently, it is possible to fuse these bodies of knowledge in a solid, powerful and flexible way. MLNs have provided a basic tool to get these together, imposing an arbitrary probability distribution over them and applying rules induced from domain knowledge (by domain experts or extracted automatically) to them to conclude new results.

## 3  The Level Set Method

In the image analysis part, we have implemented massively GPU-parallel-distributed CUDA coding of the algorithms to extract the vasculature from "raw" images, to form a "geometric model" made of B&W pixels, and to quantify the morphological traits (see [7] and [10]). The outline of the mathematical steps are as follows: Represent the image for one gray-scale frame as $f : J \to R$, and use time-index of images by Greek $\tau$ and time-dependence in the image processing algorithms by $t$. The basic idea is as follows: the typical image $f : J \to R$ is written as an outcome of an unknown convolution kernel $K$ with compact support (e.g. a discrete Gaussian), applied to the original image $\varphi$ (or a desirable model that in our case must be suitable for segmentation and midline recovery of the branched structures, and an additive noise $v$ that we propose to model as a Gaussian white noise $f = K * \varphi + v$. The automation algorithm proposes to select the kernel. In epidemiological applications we must confront derivation of a suitable model of noise that will be constructed from the imaging system outputs in lieu of the simplifying assumption above. Begin with the constrained minimization problem with appropriate norms, without further mention for simplicity.

For a class of images according to our protocol, the higher resolution allowed us to select a delta function, and use the simpler form. However, in the case of blurring or temporal loss of the automated focus in large-scale image acquisition, we must adhere to the general form below

$$\varphi = \arg\min \{ F(\varphi) := \frac{1}{2} \| K * \varphi - f \|^2_{L_2} + c \| \varphi \|_{BV} \} \tag{1}$$

Here, $c$ is a scaling constant that is the trade-off between noise and the desired image quality, such as having sharp edges in images of blood vessel branched structures, and arg-min is over all images $\varphi$, where $\varphi$ is the desired form of the image suitable for algorithms, which use an iterative regularization to recover finer scales.

An intermediate problem that we propose to solve is GPU-massive parallel implementation of methods based on the following simplified ideas (also variously appear in independent publications loc cit). Start with Chan-Wong-Kaveh-Osher et al [6] proposal for constrained optimization with $F(\varphi, K)$ as a function of joint variables $(\varphi, K)$:

$$(\varphi, K) = \arg\min_{\varphi, K} \{ F(\varphi, K) := \frac{1}{2} \| K * \varphi - f \|^2_{L_2} + c_1 \| \varphi \|_{BV} + c_2 \| K \|_{BV} \} \tag{2}$$

When we view $F$ as a one-variable functional depending only on $K$ or only $\varphi$, by fixing one, then $F$ is convex, while $F(\varphi, K)$ *fails to be* jointly convex. To overcome this difficulty, we follow Osher et al and propose to solve the two Euler-Lagrange equations (below, where $L_2$-conjugates are given by the "hat" decoration) alternatively:

$$(\varphi, K) = \arg\min_{\varphi, K}\{F(\varphi, K) := \frac{1}{2}\| K*\varphi - f \|^2_{L_2} + c_1\|\varphi\|_{BV} + c_2\|K\|_{BV}\} \tag{3}$$

$$F_\varphi(\varphi, K) = \widehat{K}*(K*\varphi - f) - c_1\nabla.\frac{\nabla\varphi}{|\nabla\varphi|} = 0$$

$$F_K(\varphi, K) = \widehat{\varphi}*(\varphi*K - f) - c_2\nabla.\frac{\nabla K}{|\nabla K|} = 0 \tag{4}$$

Fixing K first, solves for $\varphi$ from the first equation and then switch the roles of $K$ and $\varphi$, also the two equations, respectively.

In our 2010 article [7] and subsequent developments, we have developed a set of new algorithms and their object-oriented C-code for massively parallel-distributed hardware. In this ameliorated version, our research benefits from design of advanced numerical and symbolic computation algorithms by Lambe et al, and we have gained speed and efficiency that are orders of magnitude better, thus several practical challenges and bottlenecks of past approaches are overcome. Further, we have developed the codes for massively parallel-distributed platforms using the code-generation capabilities.

## 4 The Symbolic Active Contour Method

In classical implementations of the active contour algorithms by level set method, the contours around objects approximated as a zero level of the function which is the numerical solution of equation (2) [8],[9].
It is possible to replace these numerical methods by deriving an analytic model of the active contour and obtain analytic expressions for their gradients using symbolic computation.
Specifically, we use Gaussian process regression (GPR) ([10],[11]) to fit analytic functions over subsets of original image and ExprLib ([12],[13]) to do the update step of "Level Set Function" LSF by symbolic calculation of gradients. Furthermore, ExprLib has the capability to generate source code that represents these calculations. Thus we have updated our method of level sets by:
- Interpolating the image using GPR and generating C code for the model using the relevant data that was generated using ExprLib's code generating facilities,
- Using the analytic expressions generated above in ExprLib to obtain the gradients symbolically and then generating C code for those gradients,
- Incorporating the code for the contours and gradients into our procedure.

This hybrid approach, interpolating image pieces and doing symbolic computation, has following benefits in analysis of retina images:

- The accuracy of active contour segmentation would be improved because of precise calculation of derivatives.
- Measuring contour deformation in time-lapse images with analytic form of contours would be more robust and informative. We can also take the derivatives of contour deformation respect to time to find out the dynamics of the retinal vessel deformation.

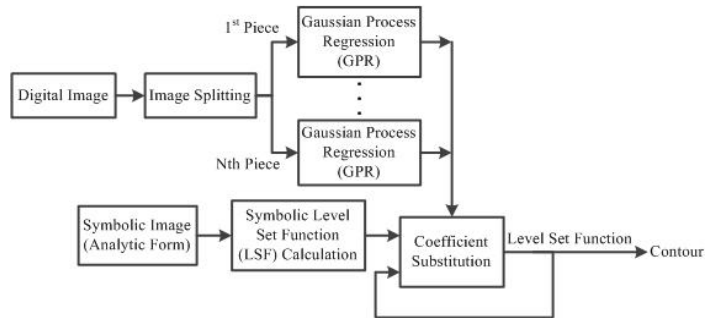

**Fig. 2.** Flow chart of the symbolic active contour method.

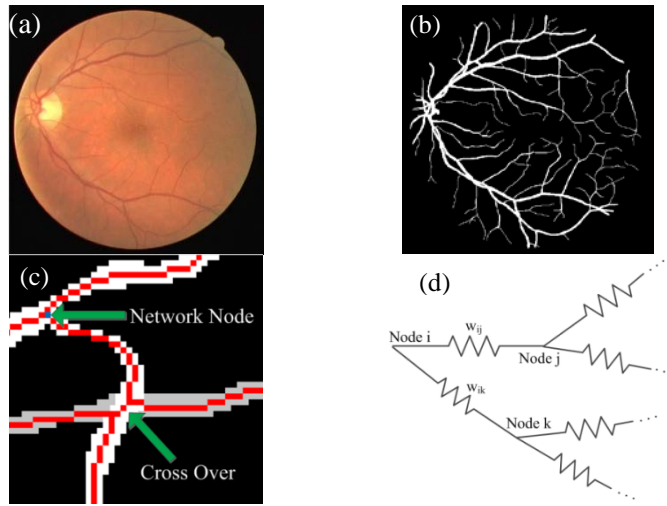

**Fig. 3.** Construction of resistance network from a raw image of retina. (a) Raw image from retina, (b) Retina in black and white, (c) Detection of network node, radius of vessels and cross overs, (d) the resistance network.

## 5   Total Effective Resistance Using Weighted Laplacians

The spectral theory of graphs has been well studied over the last few decades. While the original body of research in spectral graph theory examined adjacency matrices, more recent work has focused on the spectrum associated to the Laplacian of a graph. The Laplacian of a graph, denoted by L, forms the discrete analog of the Laplace-Beltrami operator of spectral geometry. For a simple graph G = (V, E) with |V | = n and the vertices of G labeled as {1,2,...,n}, The Laplacian L is the difference Δ − A of the degree matrix Δ (the diagonal matrix with $D_{ii} = \delta_i$) and the adjacency matrix. For a graph with non-negative edge weights $w_{ij}$, the analogue of the adjacency matrix is the matrix of weights W = ($w_{ij}$), the weighted Laplacian is LW = S − W, where S is the diagonal matrix of strengths, with elements $S_{ii} = s_i$. Lw is an object of interest because it arises as the discrete analog of the Dirichlet space metric. Symmetric functions of eigenvalues of Lw (convex or concave) are well-known invariants in the study of graph spectra. Total effective resistance, utilizes the decreasing set of eigenvalues of the weighted (normalized) Laplacian to define one such invariant as follows:

$$T = \sum_{i=2}^{n} \lambda_i^{-1} \qquad (5)$$

In analogy to the Kirchhoff circuit laws, T provides a global measure for the total resistance in a network of series-parallel connected elements [13]. Key properties of the total effective resistance T have been explored in detail [13]. For example, T is a non-increasing function of the edge weights and acts as a metric on the space of weights on the graph G. Furthermore, because T is directly related to network criticality [14], it provides a key measure of network robustness. More specifically, viewing the branching vascular system and their corresponding lengths and diameters as an analog to a resistive network, T provides a suitable global invariant for the vascular bed.

## 6  Discussion

We have outlined the blueprint for construction of a probabilistic framework to integrate various heterogeneous pieces of information using a logic network representation discussed. In this setting, descriptive data such as a health caretaker's observations, prognosis and prescriptions, are analyzed for information extraction via a representation as a set of first-order logic predicates (e.g. hasVeryLimitedEyesight(PatientX).)

Then, using a manually written or automatically extracted set of rules from various sources, a scalable MLN engine will find the most probable state upon which inference could be done. A broad variety of questions could be asked about relationships among such different pieces of information that we have acquired from various sources. Thus one acquires knowledge from information, which comes with a confidence degree that could be utilized to keep track of the environment uncertainty and changes in the complexity of the diagnostic procedure.

# 7 Conclusions

We have outlined a flexible framework for unification of various types of data which could be employed to track diseases of the retina, and lay out the basis for preventive medicine and the most effective treatments according to phenotypic traits and other characteristics in individuals within a population. Information extracted from patients medical records in the form of descriptive data, images captured over the treatment time span and different forms of numerical data could be processed in this framework using a variety of techniques which in turn could be used to track the health status of patients and study the epidemiologic traits in the population. This enables the public health systems to look for early alerts and to account for preventive measures.